\input{aipcheck}

\documentclass[
    ,final            
  ]
  {aipproc}

\layoutstyle{6x9}

\usepackage{amsfonts,amsmath,amssymb,bm,eucal}
\usepackage{graphicx,graphics}
\usepackage{epstopdf}
\usepackage{subfigure}

\begin{document}

\title{
In-medium effects on chiral condensate through holography}


\author{Floriana Giannuzzi}{
  address={Universit\`a degli Studi di Bari \& Istituto Nazionale di Fisica Nucleare, Sezione di Bari, \\
via Orabona 4,
 I-70126, Bari, Italy} 
}

\begin{abstract}
The behavior of chiral condensate in a hot and dense medium is investigated within a phenomenological approach inspired by the AdS/CFT correspondence.
We find that it decreases when temperature and/or chemical potential are increased, and it vanishes at some critical values. The locus of such critical points is also drawn and discussed. 
\end{abstract}

\maketitle


The QCD vacuum can be described in terms of some non-perturbative quantities, called condensates. 
One of these, the chiral condensate, i.e. the vacuum expectation value of the $\bar q q$ operator, is interpreted as an order parameter for the chiral transition.
In fact, at zero temperature and density, QCD (with $n_f$ massless quarks) exhibits a global symmetry, the chiral symmetry, according to which the theory is invariant under transformations of the group $U(n_f)_L\times U(n_f)_R$. 
This symmetry spontaneously breaks down to $SU(n_f)_V\times U(1)_B$, with a residual $U(1)_A$ anomalous symmetry, and $(n_f^2-1)$ Goldstone bosons appear. 
In the meanwhile, the chiral condensate becomes finite, getting the value $\sim (0.24\mbox{ GeV})^3$ \cite{colangelokho}.

Some non-perturbative approaches to QCD have shown that the chiral condensate remains finite as temperature and density increase, and it vanishes when some critical values are reached \cite{Bazavov:2009zn},
meaning that chiral symmetry is restored.
The investigation of the QCD vacuum can also be afforded by new techniques introduced by the gauge/gravity duality \cite{maldacena1}.
The AdS/QCD correspondence aims at finding a gravity theory whose lower dimensional projection contains as many as possible properties of QCD. 
The problem has been dealt with from both a top-down and a bottom-up point of view (for a review, see, e.g., \cite{Erdmenger:2007cm} and references therein).
In this respect, some phenomenological models have been introduced, in which the QCD-objects (local gauge-invariant operators) that one wants to study are translated into proper five-dimensional objects (fields) living in a anti-de Sitter (AdS) space.

In this paper the chiral condensate will be computed at finite temperature and density in the soft wall model \cite{sw}, which is one of the bottom-up holographic approaches to QCD.
Let us first define the five-dimensional model and show the rules relating it to QCD.
First, the AdS/CFT correspondence \cite{maldacena1} states that the five-dimensional space-time is a AdS$_5$ space, with metric given by 
\begin{equation}\label{metric}
ds^2=\frac{R^2}{z^2}(dt^2-d\bar x^2-dz^2)\,,
\end{equation}
where $R$ is the radius and $z$ the conformal coordinate.
Second, the holographic description suggests that  the four-dimensional gauge theory can live on the boundary of the AdS space, which is, in fact, a Minkowski space, and is defined by $z=0$. Therefore, $(t,\bar x)$ in \eqref{metric} are the four coordinates of the Minkowski space of the gauge theory.
Moreover, an equality between the partition function of the gravity theory and the generating functional of the correlators of the gauge theory has been established \cite{Witten:1998qj} 
\begin{equation}\label{adscft}
Z_5[\phi_0]=\langle  e^{\int_{\partial AdS} d^4x \, \phi_0 \CMcal{O}} \rangle\,.
\end{equation}
Eq. \eqref{adscft} also shows that the source $\phi_0$ of an operator $\CMcal{O}$ is equal to the boundary value of the field $\phi|_{z=0}=\phi_0$ dual to that operator.
This is the ultraviolet boundary condition for the field.
Finally, it is worth mentioning that the two coupling constants of the theories are related, in such a way that the supergravity limit of the theory in the AdS space (i.e., small coupling and large radius limit) corresponds to the large $N_c$ and large 't Hooft coupling regime of the gauge theory. So, if applied to QCD, the correspondence can help in studying its non-perturbative regime by studying its higher-dimensional gravity dual in the semiclassical limit, which is easier to handle and suitable for doing calculations.
However QCD has not all the features of the CFT theory described by the Maldacena conjecture.
For example, one of the main differences is that QCD is not conformal, apart from some windows of energy \cite{Brodsky:2006uqa}, so it is necessary to introduce some models which make the conjecture suitable for describing QCD. 
In particular, in the soft wall model conformal symmetry is broken by a non-dynamical dilaton field $e^{-\varphi(z)}$, $\varphi=c^2z^2$, where $c$ is a mass scale which breaks scale invariance of the theory, usually fixed by the mass of the $\rho$ meson to the value $c=0.388$ GeV \cite{sw}.\\ 
Temperature and chemical potential can be introduced in the five-dimensional theory by inserting a charged black-hole in the AdS space, and adding a U(1) gauge field $A_0$, dual to the QCD operator $\bar q\gamma_0 q$. This is known as the Reissner-Nordstrom (RN) metric
\begin{equation}
ds^2=\frac{R^2}{z^2} \left(f(z)dt^2-d\bar x^2-\frac{dz^2}{f(z)}\right) \qquad 0<z<z_h
\end{equation}
\begin{equation}\label{effe}
f(z)=1-(1+Q^2)\left(\frac{z}{z_h}\right)^4+Q^2 \left(\frac{z}{z_h}\right)^6 \qquad\quad 0\leqslant Q\leqslant \sqrt{2}\,.
\end{equation} 
$z_h$ is the outer horizon of the black hole, i.e. the lowest value of $z$ such that $f(z_h)=0$, while $Q$ is proportional to the black-hole charge; we assume
\begin{equation}\label{anot}
A_0(z)=\mu - \kappa \frac{Q^2}{z_h^3} z^2\,,
\end{equation}
with $\kappa$ a parameter, as in \cite{Colangelo:2012jy}. 
$\mu$ is the QCD chemical potential, in fact, from \eqref{anot}, we find $A_0(0)=\mu$, which is, by definition, the source of the operator $\bar q\gamma_0 q$.
Temperature and density are related to $z_h$ and $Q$: temperature is defined by
\begin{equation}
T=\frac{1}{4\pi} \left| \frac{df}{dz} \right|_{z_h} = \frac{1}{\pi z_h} \left( 1-\frac{Q^2}{2}\right)\,,
\end{equation}
while the chemical potential is obtained by the condition $ A_0(z_h)=0$
\begin{equation}
\mu=\kappa\frac{Q}{z_h}\,.
\end{equation}
It turns out that both temperature and chemical potential are inverse functions of $z_h$, so low $T$ or low $\mu$ values correspond to a larger spacetime.

Once we have defined the model, we can introduce the object we want to study, i.e. the chiral condensate. 
The QCD $\bar q q$ operator is dual to a scalar field $X(x,z)$, with mass $m_5^2=-3$ \cite{Erlich:2005qh}.
We can write $X(x,z)=(X_0(z) {\bf 1}_{n_f}+S(x,z)) e^{i\pi(x,z)}$, where $S(x,z)$ represents the fluctuations around the configuration $X_0(z)$ \cite{Colangelo:2008us} and $\pi(x,z)$ the chiral fields. 
In this description, the chiral condensate is contained in the v.e.v. $X_0$, in particular it is proportional to the coefficient of the $z^3$ term in the low-$z$ expansion.
Let us focus on the part of the action relevant for the description of its dynamics 
\begin{equation}\label{eqaction}
S=-\frac{2}{k}\int d^5x \sqrt{g}\, e^{-\varphi(z)} \left[g^{\mu\nu}\partial_\mu X_0 \partial_\nu X_0 - 3 X_0^2\right]\,,
\end{equation}
where $g$ is the determinant of the metric and $\mu,\nu$ are five-dimensional indices.
The coefficient $k$ is fixed by a comparison of the perturbative terms of the two-point correlation functions of scalar mesons in QCD and in the soft-wall model
\cite{Colangelo:2008us}.
Given the action \eqref{eqaction}, the Euler-Lagrange equation of motion for $X$ reads
\begin{equation}\label{eomfield}
 X_0''(z,z_h)- \frac{2 z^2 f(z)+4-f(z)}{z f(z)} X_0'(z,z_h)+\frac{3}{z^2 f(z)} X_0(z,z_h)=0 \,;
\end{equation}
the prime indicates a derivative with respect to $z$.

\begin{figure}[ht]
\begin{minipage}[c]{.5\textwidth}
\includegraphics[scale=0.6]{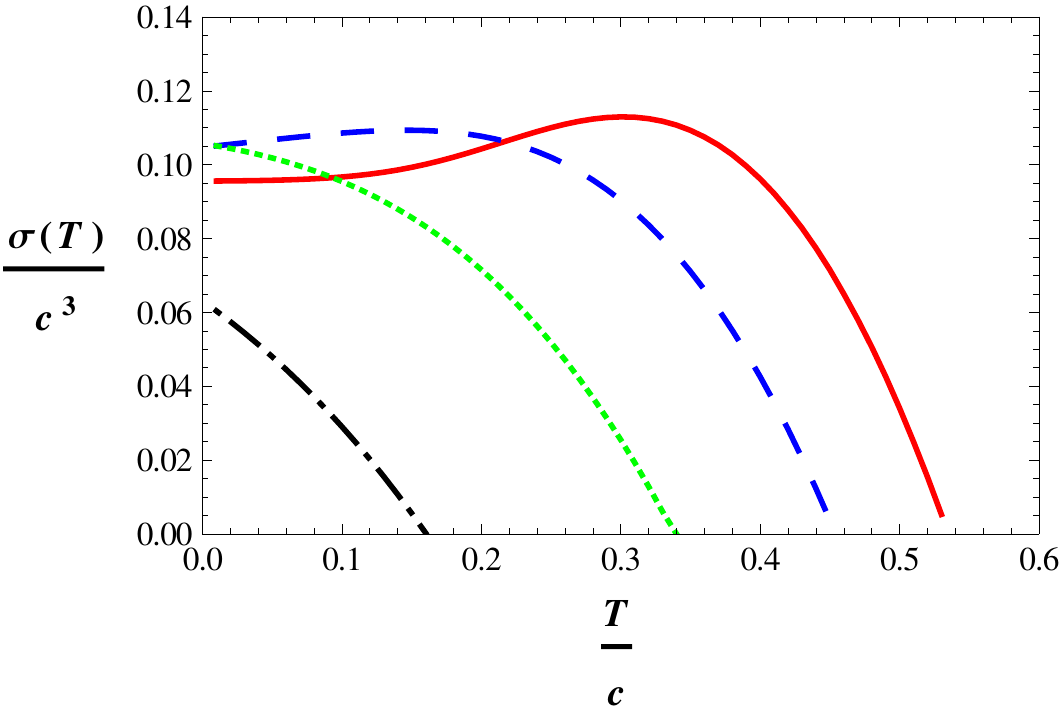}
\end{minipage}%
\begin{minipage}[c]{.5\textwidth}
\includegraphics[scale=0.6]{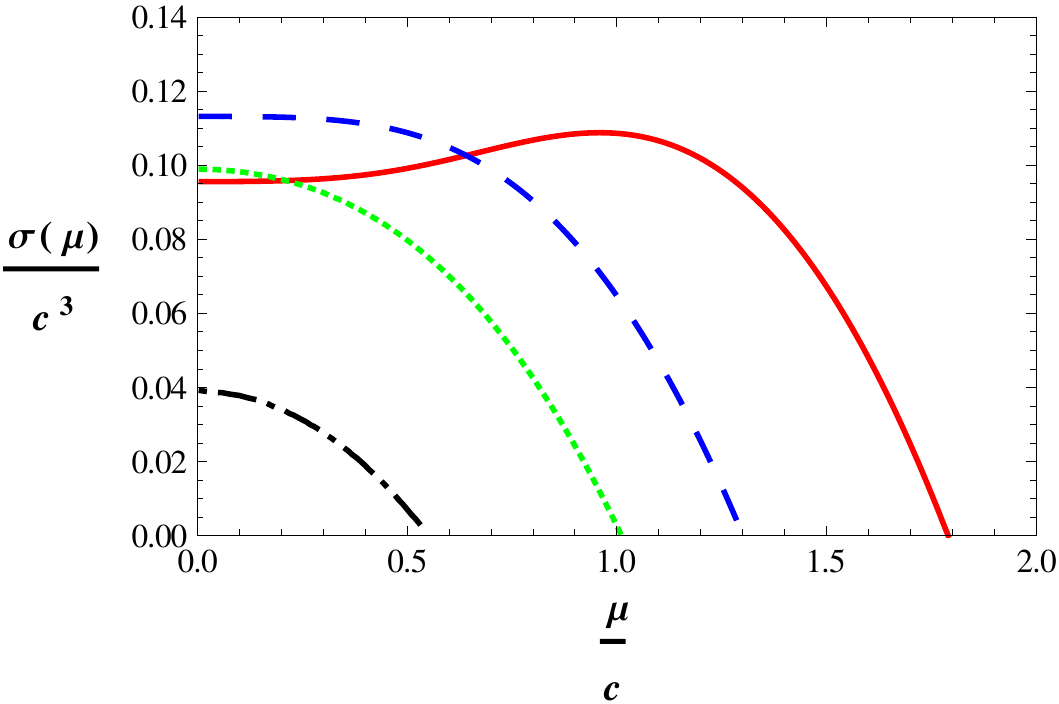}
\label{fig:sigmaparam}
\caption{Left panel: Chiral condensate versus temperature at different values of the quark chemical potential: $\mu/c=0.02$ (plain red curve), $0.8$ (blue dashed curve), $1.2$ (green dotted curve) and $1.6$ (black dot-dashed curve).
Right panel: chiral condensate versus $\mu$ at different values of $T$: $T/c=0.05$ (plain red curve), $0.3$ (blue dashed curve), $0.4$ (green dotted curve) and $0.5$ (black dot-dashed curve). $\kappa$=1 has been used.}
\end{minipage}
\end{figure}

At zero temperature and density, the metric is \eqref{metric} ($f$=1), and the solution of \eqref{eomfield} is
\begin{equation}\label{tricomi}
 X_{0}^{T=0}(z)=\frac{m_q\sqrt{\pi}}{2}\, z\, U\left(\frac{1}{2},0,z^2\right)\,,
\end{equation}
after requiring regularity in the IR ($z\to \infty$) and the UV condition $X_0/z\to m_q$ (the quark mass) at $z\to 0$.
With the identification of the chiral condensate as the coefficient of the $z^3$ term in the low-$z$ expansion of the solution, we find that it turns out to be proportional to the quark mass, at odds with what happens in QCD. 
This drawback can be overcome by modifying the model, for instance changing the dilaton field and introducing a potential term for the field $X_0$ in the action, as proposed in \cite{sw,Gherghetta:2009ac}.
However, here let us study the behavior of the chiral condensate at increasing temperature and density in the simple soft wall model, without further complications, assuming, for instance, that $m_q$ is a generic mass paramater, which fixes the mass scale of the chiral condensate.
In particular, we fix $m_q=1$.
So, we now have to solve Eq.\eqref{eomfield} with $f(z)$ in \eqref{effe}, requiring the same UV boundary condition as above, and regularity near the black-hole horizon. 
Once the solution has been numerically found, the low-$z$ expansion is performed, and, as defined above, the coefficient of the $z^3$ term is proportional to the chiral condensate\footnote{in this numerical procedure, one must remember that the low-$z$ expansion contains also a linear term and a term of order $z^3\log z$, as in \eqref{tricomi}.}.
This computation can be repeated for any $T$ and $\mu$, getting the values of the chiral condensate shown in Fig.~\ref{fig:sigmaparam}: in the left (right) panel we have obtained the behavior of the chiral condensate for some fixed values of the chemical potential (temperature) as a function of the temperature (chemical potential) \cite{Colangelo:2011sr}.
All the quantities are dimensionless, in units of $c$.

\vspace{0.3cm}

\begin{figure}[h!]
 \centering
 \includegraphics[scale=0.6]{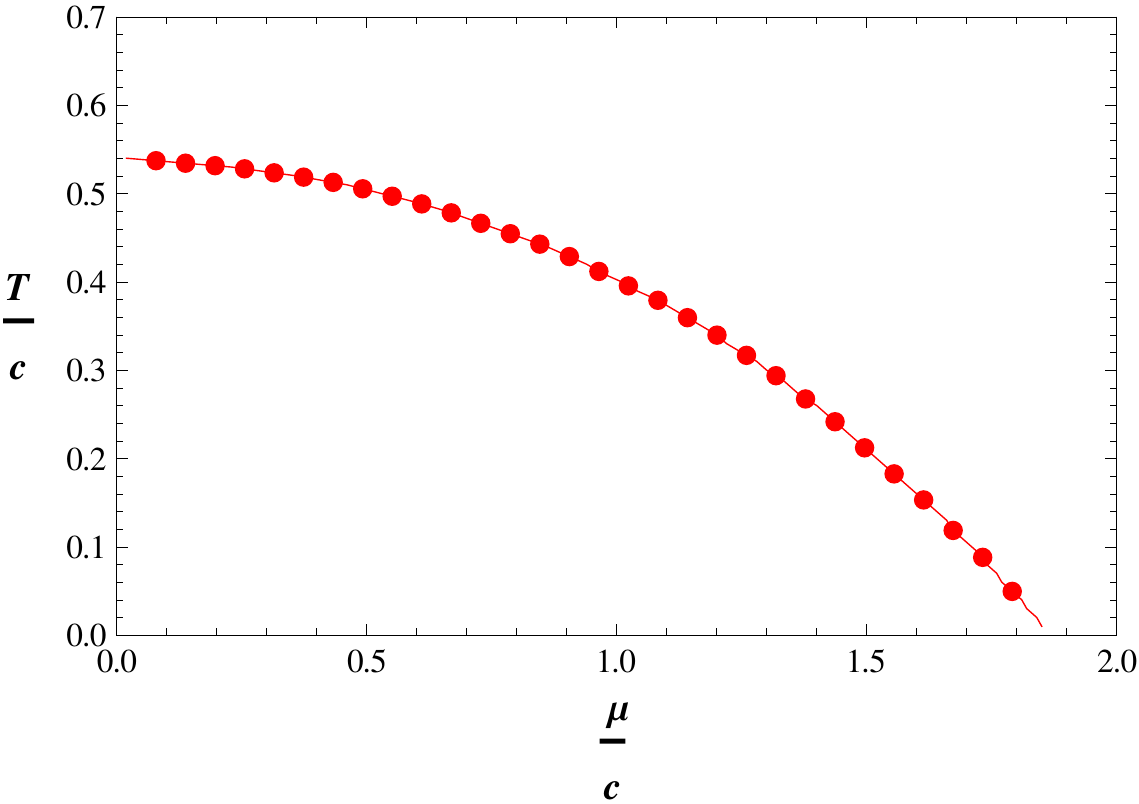}
\caption{Critical points in the $T-\mu$ plane such that $\sigma$=0. $\kappa=1$ has been fixed.}
 \label{fig:phd}
\end{figure}

We can see that the chiral condensate decreases at increasing temperature and density, and it vanishes at some critical values. This result is in agreement with calculations in different approaches, like \cite{Barducci:1993bh}. 
A QCD sum rule analysis at finite temperature and zero chemical pontential uses for the chiral condensate the profile $\sigma(T)=\sigma_0\left(1-(T/T_c)^\alpha\right)$ for temperatures close to $T_c$ \cite{Dominguez:2007ic}. The behaviour in Fig.~\ref{fig:sigmaparam} (left panel) is in agreement with \cite{Dominguez:2007ic}, as well as with the predictions of other approaches \cite{Blank:2010bz}. 
Quantitatively, we find that at $\mu=0$ the chiral condensate vanishes at $T\approx 210$ MeV, while at $T\approx 0$ the value of the chemical potential at which $\sigma$ vanishes is $\tilde\mu\sim 350$ MeV, using $\kappa=1/2$ as in \cite{Colangelo:2010pe}. 
For higher values of $T$ and $\mu$, $\sigma$ becomes negative, therefore this model can be no more reliable.
Then, one can collect the critical values $(\tilde\mu,\tilde T)$ at which $\sigma$=0, as in Fig.~\ref{fig:phd}. 
The resulting curve can be interpreted as the one dividing to phases: in the one under the curve, the $X_0$ field is different from zero, its dynamics is described by Eq. \eqref{eqaction} and it is responsible for chiral symmetry breaking; in the phase above the curve, this model is no longer reliable, and one should start with a different model, without $X_0$, in which chiral symmetry is restored. 

As a conclusion, we state that, in spite of the difficulties of the soft wall model in describing separately the explicit and spontaneous chiral symmetry breaking, this model is able to reproduce a decreasing behavior of the chiral condensate at increasing temperature and density, and the restoration of chiral symmetry. Moreover, the locus of points in the $\mu,T$ plane in which $\sigma$=0 is a reliable picture of the chiral transition in the QCD phase diagram.


\begin{theacknowledgments}
I would like to thank P. Colangelo and S. Nicotri for collaboration. 
\end{theacknowledgments}



\bibliographystyle{aipproc}   

\end{document}